\documentclass[manuscript]{aastex631}
\usepackage{amsmath} 
\usepackage{amsbsy}

\received{xx}
\revised{xx}
\accepted{xx}
\submitjournal{ApJ}

\shorttitle{Interplanetary Rotation of 2021 December 4 CME}
\shortauthors{Ma et al.}

\begin{document}

\title{Interplanetary Rotation of 2021 December 4 Coronal Mass Ejection on its Journey to Mars}

\correspondingauthor{Liping Yang, Fang Shen}
\email{lpyang@swl.ac.cn, fshen@swl.ac.cn}

\author{Mengxuan Ma}
\affiliation{SIGMA Weather Group, State Key Laboratory for Space Weather, National Space Science Center, Chinese Academy of Sciences, Beijing 100190, People's Republic of China}
\affiliation{College of Earth and Planetary Sciences, University of Chinese Academy of Sciences, Beijing 100049, People's Republic of China}

\author[0000-0003-4716-2958]{Liping Yang}
\affiliation{SIGMA Weather Group, State Key Laboratory for Space Weather, National Space Science Center, Chinese Academy of Sciences, Beijing 100190, People's Republic of China}
 
\author{Fang Shen}
\affiliation{SIGMA Weather Group, State Key Laboratory for Space Weather, National Space Science Center, Chinese Academy of Sciences, Beijing 100190, People's Republic of China}
\affiliation{College of Earth and Planetary Sciences, University of Chinese Academy of Sciences, Beijing 100049, People's Republic of China}

\author{Chenglong Shen}
\affiliation{Deep Space Exploration Laboratory/School of Earth and Space Sciences, University of Science and Technology of China, Hefei 230026, People's Republic of China}
\affiliation{CAS Center for Excellence in Comparative Planetology, University of Science and Technology of China, Hefei 230026, People's Republic of China}

\author[0000-0001-9315-4487]{Yutian Chi}
\affiliation{Institute of Deep Space Sciences, Deep Space Exploration Laboratory, Hefei 230026, People's Republic of China}

\author{Yuming Wang}
\affiliation{Deep Space Exploration Laboratory/School of Earth and Space Sciences, University of Science and Technology of China, Hefei 230026, People's Republic of China}
\affiliation{CAS Center for Excellence in Comparative Planetology, University of Science and Technology of China, Hefei 230026, People's Republic of China}
\affiliation{Anhui Mengcheng Geophysics National Observation and Research Station, University of Science and Technology of China, Mengcheng, Anhui, People ’ s Republic
of China}

\author{Yufen Zhou}
\affiliation{SIGMA Weather Group, State Key Laboratory for Space Weather, National Space Science Center, Chinese Academy of Sciences, Beijing 100190, People's Republic of China}

\author{Man Zhang}
\affiliation{SIGMA Weather Group, State Key Laboratory for Space Weather, National Space Science Center, Chinese Academy of Sciences, Beijing 100190, People's Republic of China}

\author{Daniel Heyner}
\affiliation{Institut fr Geophysik und extraterrestrische Physik, Technische Universitt Braunschweig, Braunschweig, Germany}

\author{Uli Auster}
\affiliation{Institut fr Geophysik und extraterrestrische Physik, Technische Universitt Braunschweig, Braunschweig, Germany}
\author{Ingo Richter}
\affiliation{Institut fr Geophysik und extraterrestrische Physik, Technische Universitt Braunschweig, Braunschweig, Germany}

\author{Sanchez-Cano, Beatriz}
\affiliation{School of Physics and Astronomy, University of Leicester, Leicester, UK}



\begin{abstract}

The magnetic orientation of coronal mass ejections (CMEs) is of great importance to understand their space weather effects.
Although many evidences suggest that
CMEs can undergo significant rotation during the early phases of evolution in the solar corona,
there are few reports that CMEs rotate in the interplanetary space.
In this work, we use multi-spacecraft observations and a numerical simulation starting from the lower corona close to the solar surface to understand
the CME event on 2021 December 4, with an emphatic investigation of its rotation.
This event is observed as a partial halo CME from the back
side of the Sun by coronagraphs, and reaches the BepiColombo spacecraft and the MAVEN/Tianwen-1 as a magnetic flux rope-like structure.
The simulation discloses that in  the solar corona the CME is approximately a translational motion, while
the interplanetary propagation process evidences a gradual change of axis orientation of
the CME's flux rope-like structure.
It is also found that the downside and the right flank of the CME moves with the fast solar wind, and the upside does
in the slow-speed stream. The different parts of the CME with different speeds generate the nonidentical displacements of its magnetic structure, resulting in the rotation of the CME in the interplanetary space.
Furthermore, at the right flank of the CME exists a corotating interaction region (CIR), which makes the orientation of the CME alter, and also deviates from its route due to the CME.
These results provide new insight on interpreting CMEs' dynamics and structures during their travelling through the heliosphere.

\end{abstract}



\section{Introduction} \label{sec:intro}

 Coronal mass ejections (CMEs) contain a large amount of energetic plasma and magnetic fields
that frequently erupt from the low solar corona and advance into the interplanetary space \citep{Chen2011a, Zhang2021}.
The evolution of CMEs involves many physical processes, such as magnetic reconnection, plasma heating,
particle acceleration \citep{Manchester2017, Luhmann2020}. By the time CMEs' interplanetary manifestations (ICMEs) are directed at Earth and planets, they can induce extreme space weather in Earth's and planetary environments \citep{Gosling1993, Feng2013, Shen2017, Gopalswamy2022}. It is therefore of great interest to understand the dynamic evolution of CMEs/ICMEs for the study of fundamental
plasma processes as well as for the development of space weather forecast models.

With the development of space exploration technology and the increase of space and planetary
missions in the heliosphere, a combination of  remote-sensing observations from multiple perspectives and in situ measurements at different locations allow a deep investigation into the morphology and kinematic properties of CMEs/ICMEs \cite[e.g.][]{Burlaga1982, Liu2008a, Liu2008b, Manchester2017, Wang2018a, Lugaz2020, Chi2021, Davies2022, Xu2022, Scolini2022, Zimbardo2023, Palmerio2024}.
In particular, the solar wind parameters monitored by Mars Express (MEX) spacecraft since 2003 \citep{Barabash2006}, Mars Atmosphere and Volatile EvolutioN (MAVEN) since 2014 \citep{Jakosky2015}, and Tianwen-1 since 2021 \citep{Wan2020} provide us with the opportunity to
follow CMEs from the Sun to Mars and to feature their space weather responses at the planets \cite[e.g.][]{Crider2005, Mostl2015, Jakosky2015Sci, Luhmann2017, Lee2017, Xu2018, Zhao2021, Palmerio2022, Chi2023a, Chi2023b}. Unlike the Earth, Mars lacks global intrinsic magnetic fields.
Fields from induced ionospheric currents and localized crustal fields are not strong enough to
divert ICMEs, leading to the direct interactions between the Martian atmosphere/ionosphere and ICMEs.
Studying the propagation and dynamics of CMEs from the Sun to Mars is vital to understand
the evolution of the Martian environment.

Many CMEs tend to have a coherent magnetic structure that can be well described
by a magnetic flux rope. As CMEs depart away from the Sun and enter into the interplanetary space, they may experience rotation, deflection, and distortion as a result of their interactions with the surrounding solar wind and/or themselves' dynamics \citep{Kilpua2017, Shen2022}. The rotation of CMEs/ICMEs is an important topic as the magnetic orientation of ICMEs is one of the key factors to trigger space weather events.
From the orientation of the eruptive flux ropes, CME rotation is evidenced to occur at their initiation in the low solar atmosphere as a result of the helical kink instability \citep{Torok2010}, the torque exerted by an external shear field \citep{Isenberg2007}, magnetic reconnection with the ambient field \citep{Thompson2011}, and so on.
By fitting CMEs with elliptical cones models or the Graduated Cylindrical Shell (GCS) model in the coronagraph field of view as a
function of time, the rotation of CMEs was deduced in the middle and outer corona \citep{Krall2006, Yurchyshyn2007, Yurchyshyn2008, Lynch2010, Vourlidas2011, Chinchilla2012, Kay2017}. The rotation can be driven by the gradients of the large-scale magnetic fields or the disconnection of one
flux-rope footpoint. \cite{Isavnin2013} estimate the flux rope orientation
first in the near-Sun regions using coronagraph images of CMEs and then in-situ
using Grad-Shafranov reconstruction of the magnetic cloud, and find evidence of rotation on
the travel of CMEs from the Sun to 1 AU. According to \citet{Kay2023}, the differential drag forces at the flanks of CMEs determine a torque about the CMEs' nose, which may cause a rotation of CMEs in the interplanetary space. However, CME's rotation
is less likely to happen at large distances from the Sun due to the weak ambient magnetic field \citep{Lynch2009, Isavnin2013}.

As an efficient and important way, magnetohydrodynamic (MHD) numerical simulations are also applied to study the rotations of CMEs during their near-Sun dynamics \citep[e.g.,][]{Lynch2009, Cohen2010, Shiota2010, Kliem2012, Maharana2023, Zhang2024}.
By numerical simulations of CME initiation via the magnetic breakout model, \cite{Lynch2009} calculate the rotation of erupting
flux ropes, and find that the right-handed flux rope rotates clockwise and the left-handed flux rope rotates
counterclockwise when propagate through the low corona (from 2 to 4 $R_s$). The rotation is ascribed primarily
to the evolution of the Lorentz force during the initial expansion of the core magnetic field.
Based on a global MHD numerical study of a CME like the 12 May 1997 CME event, \cite{Cohen2010} point out that during the stage of the eruption, the
simulated flux rope rotates to achieve an orientation that conforms to the interplanetary flux rope observed at 1 AU
as a result of the interaction between the flux rope and the ambient fields via magnetic reconnection.
\cite{Maharana2023}  perform a  MHD data-driven simulation of two successive CMEs in the heliosphere with EUropean Heliosphere FORecasting Information Asset (EUHFORIA), suggesting that a significant rotation of CMEs occur in the low corona (i.e. within 0.1 au) in order to explain the observed magnetic field profile at 1 au.

In this work, we combine multi-spacecraft observations (e.g., SOHO, STEREO-A, BepiColombo, MAVEN, Tianwen-1) with a MHD numerical simulation to
investigate the rotation of 2021 December 4 CME in the interplanetary space.
As a result of expansion, this CME experiences large velocity
differences associated with the ambient solar wind, which means that some parts of  the CME have fast speed, while some parts of  the CME are with low speed. This uneven motion across the CME leads to the uneven displacements of its magnetic fields after a time interval.
As a result, the axial direction of the magnetic flux rope-like structure of the CME changes.
The paper is organized as follows:
in Section 2, we briefly describe the observation of 2021 December 4 CME as well as the setup of our numerical simulation.
In Section 3, we present observation and simulated results, and in Section 4, summary and discussion are given.

\section{Observations and Methods}

\subsection{The 2021 December 4 Event}
On 2021 December 4 at 16:00 Eastern Time (ET), a CME erupts from the back
side of the Sun relative to the Earth, with the eruption direction having an angle of approximately $145^\circ$ relative to the Earth-Sun line \citep{Chi2023b}. It appears as a halo or partial halo CME in the remote sensing observations of the Sun from the C2/C3 coronagraph observations of the SOlar and Heliospheric Observatory (SOHO)/the Large Angle and Spectrometric
COronagraph (LASCO). As the Solar TErrestrial Relations Observatory Ahead (STEREO-A) was located at $35.5^\circ$  west of the Earth On 2021 December 4, STEREO-A/COR2 also observe this CME as a back-side partial halo CME. However, the relative
positions of the CME and STEREO-A make the heliospheric imagers on board STEREO-A not follow the CME.

At the time of the CME eruption, the BepiColombo is located at a distance of 0.67 AU from the Sun, with an angle of $134^\circ$ relative to the Earth-Sun line, while MAVEN and Tianwen-1 are at a distance of 1.67 AU from the Sun and $16^\circ$ west of the BepiColombo.
As a result of the favorable locations of the BepiColombo, MAVEN and Tianwen-1, the CME are first detected by the BepiColombo on December 6, and then by MAVEN and Tianwen-1 On December 10 \citep{Chi2023a, Chi2023b}. These in situ observations show that
the CME features an increased magnetic field and smoothly changing magnetic field directions, meaning that it has a flux rope-like structure.

\subsection{Simulation Setup}
\begin{figure}[htbp]
   \begin{center}
   \begin{tabular}{c}
     \includegraphics[width = 5 in]{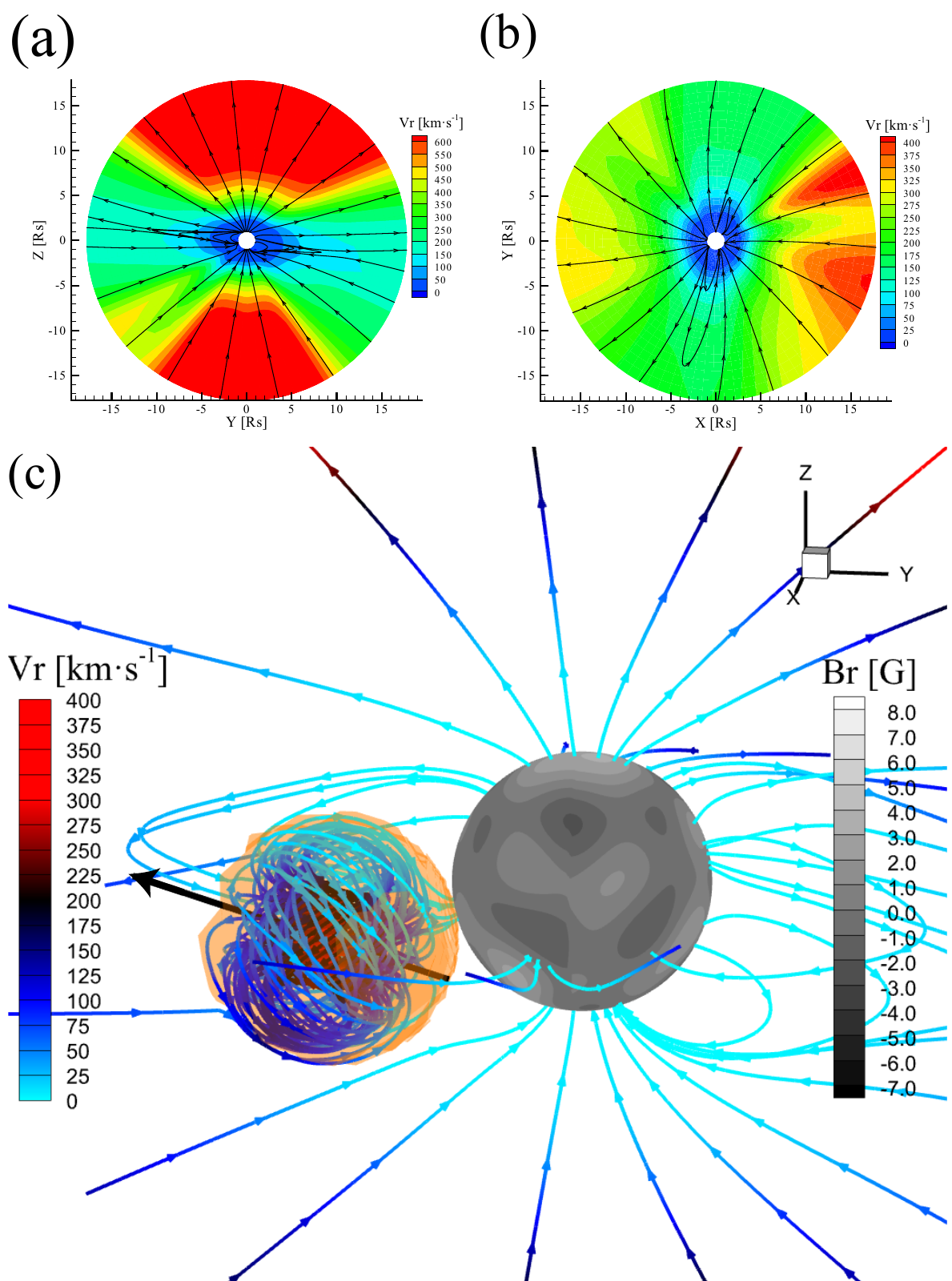}   \\
  \end{tabular}
   \end{center}
\caption{Panels (a) and (b) show the calculated quasi-steady solar-wind solution for the radial speed $Vr$ and the magnetic field lines in the equatorial plane (a) and in the meridional plane (b) from 1-18 Rs.
Panels (c) shows 3D view of the initialized spheromak CME and the calculated quasi-steady corona from 1-5 Rs. Magnetic field lines are drawn as solid color lines with the color denoting the radial speed $Vr$. The color contours show the distribution of $Br$ on the solar surface.
The orange transparent 3D surface marks the contour of the $\rho_c= 0.01$ surface.
The black line with an arrow displays the direction of the symmetry axis of the spheromak CME.}\label{fig1}
\end{figure}

The 3D adaptive mesh refinement (AMR) solar-interplanetary space-time conservation element and solution element (SIP-CESE) MHD solar wind model used in this paper is detailed in
\cite{Feng2010, Feng2012, Feng2015}, \cite{Yang2012}, and \cite{Feng2020a}. This model is also implemented to study the 3D morphology and kinematics of the 2017
September 10 CME-driven shock from the Sun to Earth \citep{Yang2021} as well as the distortion of the 2021 October 9 CME from an
ellipsoid to a concave shape \citep{Yang2023}.
To study the rotation of the 2021 December 4 CME from the solar corona to Mars, we first simulate the ambient solar wind in which the CME propagates.
We use the Parker's hydrodynamic isothermal solar wind solution \citep{Parker1963} to obtain the initial distributions of plasma
density, $\rho$, gas pressure $p$, and plasma velocity $\mathbf{u}$. The temperature and the number density on the solar surface
are set to be $1.45\times10^6$ K and $2\times10^8$ cm$^{-3}$, respectively. For the initial magnetic field,
we utilize the Global Oscillation Network Group (GONG) magnetogram of Carrington Rotation (CR) 2251 to get a potential field as input.
The inner boundary at 1 Rs is fixed for simplicity and the outer boundary at 500 Rs is dealt with equivalent extrapolation.
After that, we achieve a quasi-steady state of the solar wind in CR 2251, which shows the closed-field streamers with the low-speed plasma and the polar open fields where the high-speed solar wind is located (see Figure \ref{fig1}a,b).

\begin{table}[ht] 
\caption{The parameters derived from the GCS fitting}
\begin{tabular}{ccccccc}
\hline
\hline
latitude & longitude & half angular width & speed & tilt angle & aspect ratio\\
\hline
 $-18.0^\circ$ & $150.0^\circ$ & $45.6^\circ$ (between legs) &761.2 km s$^{-1}$ & $-70.0^\circ$ & 0.43(25.5$^\circ$) \\
\hline
\end{tabular}
\end{table}

A linear force-free spheromak \citep{Kataoka2009, Zhou2014, Scolini2019, Koehn2022, Yang2021, Yang2023} is still used here to initialize the 2021 December 4 CME with the aim to capture its flux rope-like structure. The spheromak CME is overlaid onto the simulated background solar wind (see Figure \ref{fig1}c), and is confined by the parameters of the CME obtained from the graduated cylindrical shell (GCS) fitting \citep{Thernisien2011} to nearly simultaneous images from SOHO/LASCO and STEREO-A/COR2 \citep{Chi2023b}. The parameters derived from the GCS fitting are shown in Table 1. We use a passive tracer with value greater than 0 ($\rho_c>0$ ) to tag the CME \citep{Yang2023}. The orange $\rho_c= 0.01$ surface and the distribution of magnetic field lines in the Figure \ref{fig1}c clearly display that the spheromak rope is detached from the Sun,  which can avoid the angular momentum supplied through anchor foot points.
We also delicately tune the parameters of the spheromak CME to make simulation results match with white light observations and in situ
measurements. The time when the spheromak CME is superposed is set as t = 0 h.

It should be mentioned that the tilting instability of the
spheromak can be triggered if the symmetry axis of the spheromak is at an angle with respect to the direction of the ambient field.
The instability generates a rotation and tilting of the spheromak \citep{Shiota2010, Asvestari2022, Sarkar2024}.
In our simulation, we set the symmetry axis of the spheromak nearly aligning with the ambient field  (see Figure \ref{fig1}c) to avoid the occurring of the tilting instability.

\section{Results}

\begin{figure}[htbp]
   \begin{center}
   \begin{tabular}{c}
     \includegraphics[width = 6 in]{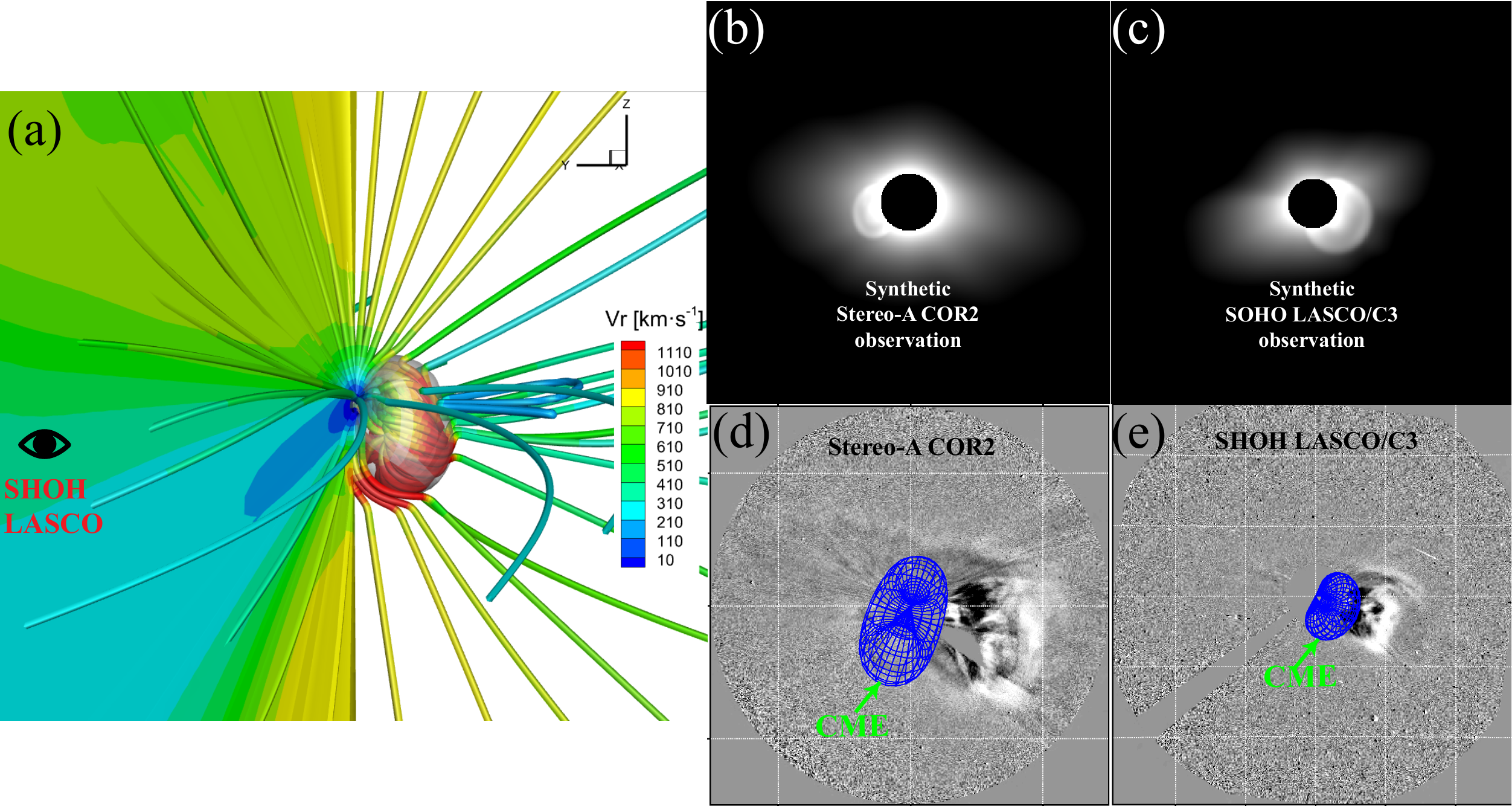}   \\
  \end{tabular}
   \end{center}
\caption{Panel (a) shows 3D magnetic flux-rope structure of the simulated CME at t = 1 h. The black eye shows the observation direction of SOHO/LASCO. Panels (b) and (c) show the synthetic white-light images of STEREO A/COR2 (b) and SOHO/LASCO C3 (c), which accord with real white-light observations by STEREO A/COR2 (d) and SOHO/LASCO C3 (e), respectively. Panel(d) and (e) show the comparison of GCS-fitted CME and real white-light observations by STEREO A/COR2 (d) and SOHO/LASCO C3 (e).}\label{fig2}
\end{figure}
Coronal polarized brightness observations are often employed to diagnose eruption positions and propagation directions of CMEs.
Based on the 3D MHD simulation data, we first synthesize the coronal polarized brightness in the views of
SOHO/LASCO C3 and STEREO-A/COR2 to check whether the simulation capture the characteristics of the CME detected in the real observations of the coronal polarized brightness.
Figure \ref{fig2}a presents the relative positions of SOHO and the simulated CME, which shows that
SOHO/LASCO observes the CME from a back view.
In Figure \ref{fig2}b, where the synthetic image of STEREO A/COR2 is displayed,
the simulated CME appears a partial halo, whose position and overall shape are comparable to
those of the CME in the white light image observed by STEREO A/COR2 (Figure \ref{fig2}d).
It is clear from Figure \ref{fig2}c,e that a qualitative agreement of the CME's positions
exists between the synthetic and real images of SOHO/LASCO C3.

\begin{figure}[htbp]
   \begin{center}
   \begin{tabular}{c}
     \includegraphics[width = 6 in]{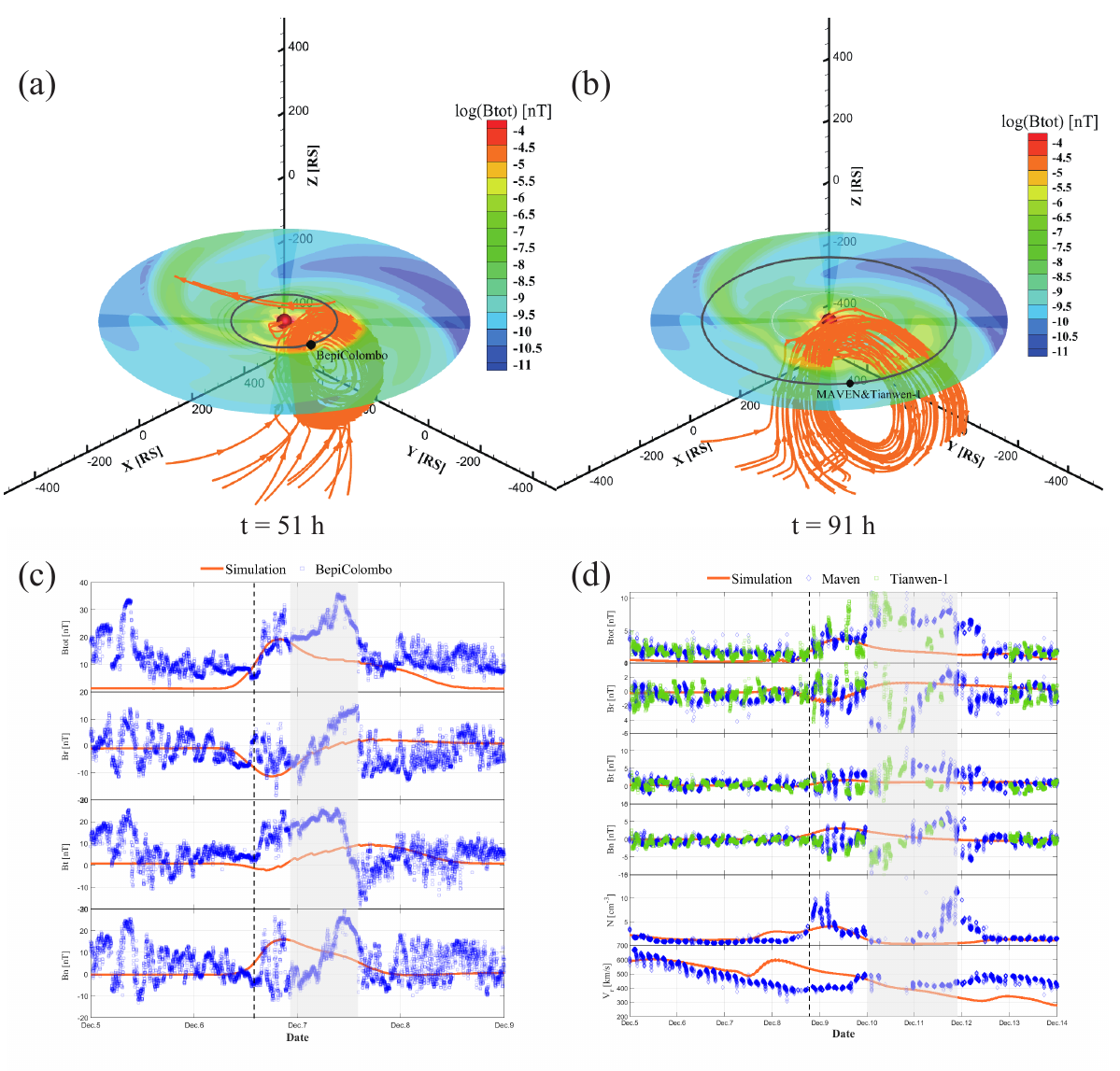}   \\
  \end{tabular}
   \end{center}
\caption{Panels (a) and (b) show 3D view of the simulated CME in interplanetary space at t = 51 h (a) and t = 91 h (b), respectively, where the slice is on the equatorial plane from 20 to 500 Rs, with the colors denoting the magnetic field strength ($B_\textrm{tot}$). The orange lines represent magnetic field lines, and the gray lines show the trajectories of the BepiColombo and MAVEN/Tianwen-1.
Panel (c) shows, from top to bottom, the magnetic field strength $B_\textrm{tot}$ and the magnetic field vector ($Br$, $Bt$, and $Bn$) in RTN coordinates measured by BepiColombo (blue squares) and reproduced by our simulation (orange lines). The vertical line and shaded area indicate the arrival of the shock and interval of the CME at BepiColombo, respectively. Panel (d) shows, from top to bottom, the magnetic field strength $B_\textrm{tot}$, the magnetic field vector ($Br$, $Bt$, and $Bn$) in RTN coordinates, the number density ($N$), and the bulk velocity ($V$) measured by MAVEN (blue squares) and reproduced by our simulation (orange lines). The vertical line and shaded area represent the shock caused by the CME and interval corresponding to the CME determined from the data of MAVEN, respectively.}\label{fig3}
\end{figure}

From the magnetic field measurements \citep{Glassmeier2010, Heyner2021} by the BepiColombo, we approximate the propagation time of the CME
from its eruption to its arrival at the BepiColombo, which is about 48 hours.
The entire transit process of the CME lasts for about 24 hours in the BepiColombo (Figure \ref{fig3}c).
The magnetic field and plasma measurements near Mars show that after 96 hours, the shock caused by the CME arrives, with the entire crossing process of the CME persisting about 4 days (Figure \ref{fig3}d). We present in Figure \ref{fig3}a,b the 3D flux rope-like structure of the simulated CME as well its distributions on the equatorial plane when the simulated CME reaches the BepiColombo and MAVEN/Tianwen-1. It can be seen that both the BepiColombo and MAVEN/Tianwen-1 pass through the up part of the CME. Although the simulation reproduces well the arrival and the magnetic field strength of the shock driven by the CME both at the BepiColombo and MAVEN/Tianwen-1, the magnetic fields at the interval of the simulated CME are smaller than those probed by the BepiColombo and MAVEN/Tianwen-1 (Figure \ref{fig3}c and d).

\begin{figure}[htbp]
   \begin{center}
   \begin{tabular}{c}
     \includegraphics[width = 6 in]{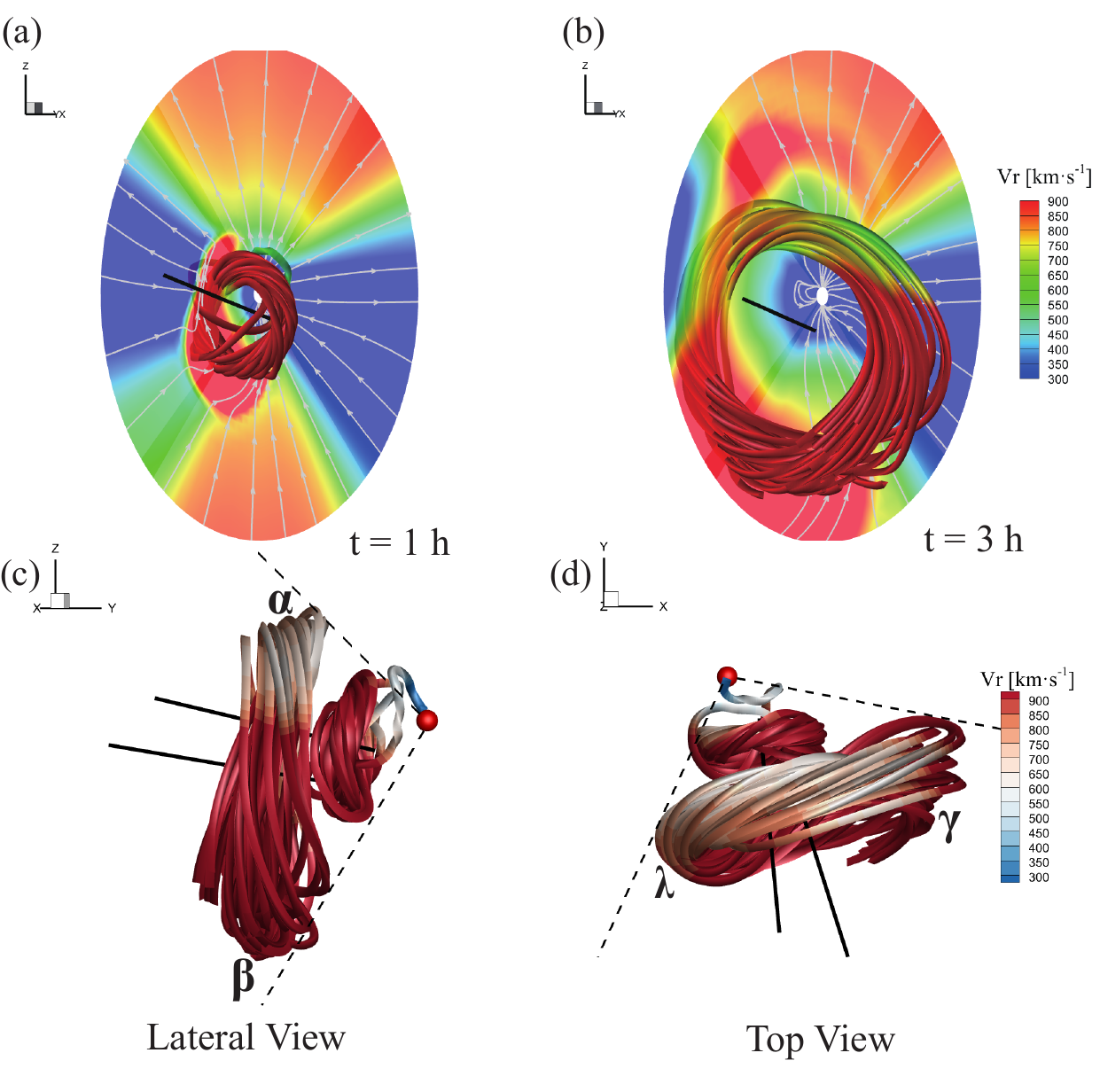}   \\
  \end{tabular}
   \end{center}
\caption{Panels (a) and (b) show 3D magnetic flux-rope structure of the simulated CME in the solar corona at $t=1$ h (a) and $t=3$ h (b), respectively, where the slice is on the plane of $x=0$ Rs with the radial range between 1 and 20 Rs. The colored lines denote magnetic field lines, and the colors represent the radial velocity ($V_r$). The black lines denote orientation of the symmetric axis of the spheromak CME.
Panels (c) and (d) show the magnetic flux ropes at $t=1$ h and $t=3$ h together at the lateral (c) and top (d) views, respectively. $\alpha$, $\beta$,  $\lambda$ and $\gamma$ mark the top,  bottom, left, and right parts of the flux rope, respectively.  The red sphere represents the Sun, and the black dashed lines are radial lines.}\label{fig4}
\end{figure}

\begin{figure}[htbp]
   \begin{center}
   \begin{tabular}{c}
     \includegraphics[width = 6 in]{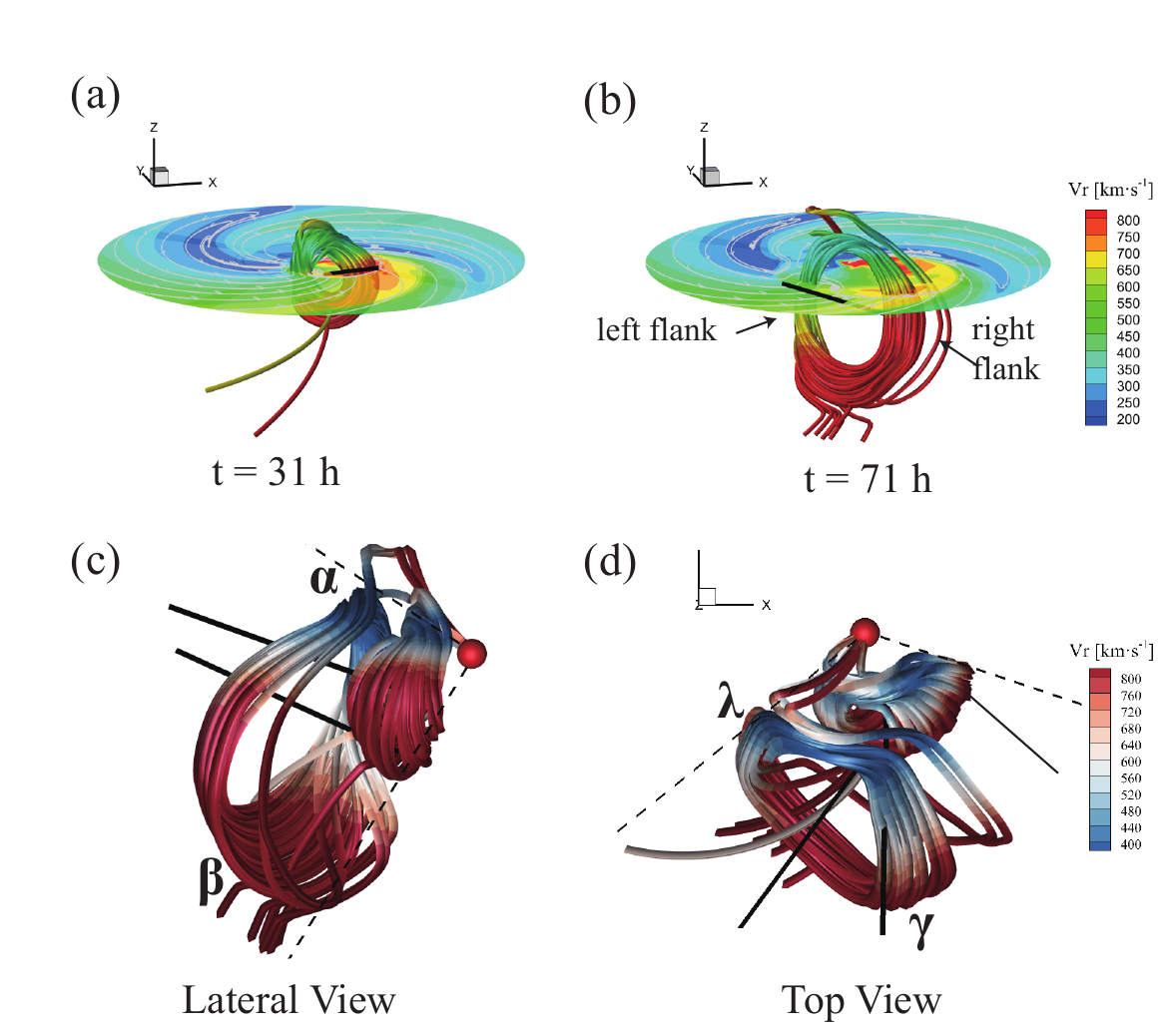}   \\
  \end{tabular}
   \end{center}
\caption{Panels (a) and (b) show 3D magnetic flux-rope structure of the simulated CME in the interplanetary space at $t=31$ h (a) and $t=71$ h (b), respectively, where the slice is on the plane of $z=-20$ Rs with the radial range between 20 and 500 Rs. The colored rods represent magnetic field lines, and the colors denote the radial velocity ($V_r$). The black lines denote orientation of the symmetric axis of the spheromak CME.
Panels (c) and (d) show the magnetic flux ropes at $t=31$ h and $t=71$ h together at the lateral (c) and top (d) views, respectively. 
$\alpha$, $\beta$,  $\lambda$ and $\gamma$ mark the top,  bottom, left, and right parts of the flux rope, respectively.  The red sphere represents the sphere with a  radius of 20 Rs, and the black dashed lines are radial lines.}\label{fig5}
\end{figure}

To study the rotation occurring during the propagation of this CME, we plot the 3D magnetic flux-rope structure of the CME from the solar corona to the interplanetary space. Figure \ref{fig4} illustrates its propagation in the solar corona (1-20 Rs), and Figure \ref{fig5} shows the interplanetary propagation process (20-500 Rs). To facilitate the judgement of the CME's rotation, we average the binormal of the magnetic field lines belonging to the CME as orientation of the symmetric axis of the spheromak CME, which is shown by the black lines in Figures \ref{fig1}, \ref{fig4} and \ref{fig5}. We use change in orientation of the symmetric axis of the spheromak CME to detect CME's rotation. According to the work by \cite{Asvestari2022}, the direction of the symmetric axis is correlated with the direction of the toroidal field axis. In our work, the CME rotation means the change of the orientation of the symmetric axis of the spheromak flux rope. The rotational motion of the CME is focused on here as differential displacements at the different parts of the spheromak flux rope after a time interval.

During the propagation in the solar corona, the CME crosses a distance of 20 Rs in just 3 hours due to its fast speed.
Figure \ref{fig4}a and b display the 3D distribution of the magnetic flux-rope structure of the CME as well as the direction of its symmetric axis at $t=1$ h and $t=3$ h, respectively, from which ones observe no significant variations of the symmetric axis's direction in the solar corona.
Figure \ref{fig4}c and d plot the magnetic flux-rope structures of the CME at $t=1$ h and $t=3$ h together, which showcase that from $t=1$ h to $t=3$ h, the distance between the top parts ($\alpha$) of the flux-rope structure is almost equal to that of the bottom parts ($\beta$), and the distance between the left parts ($\lambda$) approaches to that between the right parts ($\gamma$), which means that
the motion of the magnetic flux-rope structure of the CME is approximately translational in the solar corona.

Figure \ref{fig4} further reveals that the asymmetric CME expansion is formed in the corona as a result of the interactions with the coronal background structures. From $t=1$ h to $t=3$ h, the $\alpha$ part of the flux-rope structure is confined in the low-latitude domain with the low-speed solar wind, while the $\beta$ part of the flux-rope structure bulges into the high-latitude region where the fast solar wind is located.

As the CME enters into the interplanetary space, the speed of the CME tends to be consistent with that of the background solar wind, resulting in velocity differences between the different parts of the flux rope.
Figure \ref{fig5}a and b show that the $\alpha$ part of the flux-rope structure decelerates to about 450 km s$^{-1}$, while the $\beta$ part of the flux-rope structure keeps moving at a high speed of about 800 km s$^{-1}$. 
It can also be noted in Figure \ref{fig5}a and b that the $\gamma$ part of the flux rope is located at a high-speed stream of the background solar wind, while the $\lambda$ part seats near a low-speed stream of the background solar wind. The average speed on the $\gamma$ part of the flux-rope structure is about 200 km s$^{-1}$ higher than that on the $\lambda$ part. These speed differences across the flux rope can influence its morphology.
 
From the 3D distribution of the the magnetic flux-rope structure as well as its symmetric axis's direction in Figure \ref{fig5}a and b, it can be found that the interplanetary propagation process of the CME exhibits a significant rotation.  From $t=31$ h to $t=71$ h, the symmetric axis's direction deviates by approximately $40^\circ$. From Figure \ref{fig5}c,  which contrasts the morphology of the magnetic flux-rope structure of the CME at $t=31$ h to that at $t=71$ h at the lateral view,
it can be found that from $t=31$ h to $t=71$ h, the \text{$\beta$} part of the flux-rope structure with the high speed travels longer distance than the \text{$\alpha$} part with the slow speed. In the same way, from $t=31$ h to $t=71$ h the separation of the \text{$\gamma$} parts of the CME is more pronounced than the separation of the \text{$\lambda$} parts, which can be seen in Figure \ref{fig5}d. As the magnetic fields are frozen with the plasma, the uneven motion across the CME generates the  nonidentical displacements of its magnetic field lines after a time interval, which makes the axial direction of the flux-rope structure change. As a result, the rotation of the CME takes place.

\begin{figure}[htbp]
   \begin{center}
   \begin{tabular}{c}
     \includegraphics[width = 6 in]{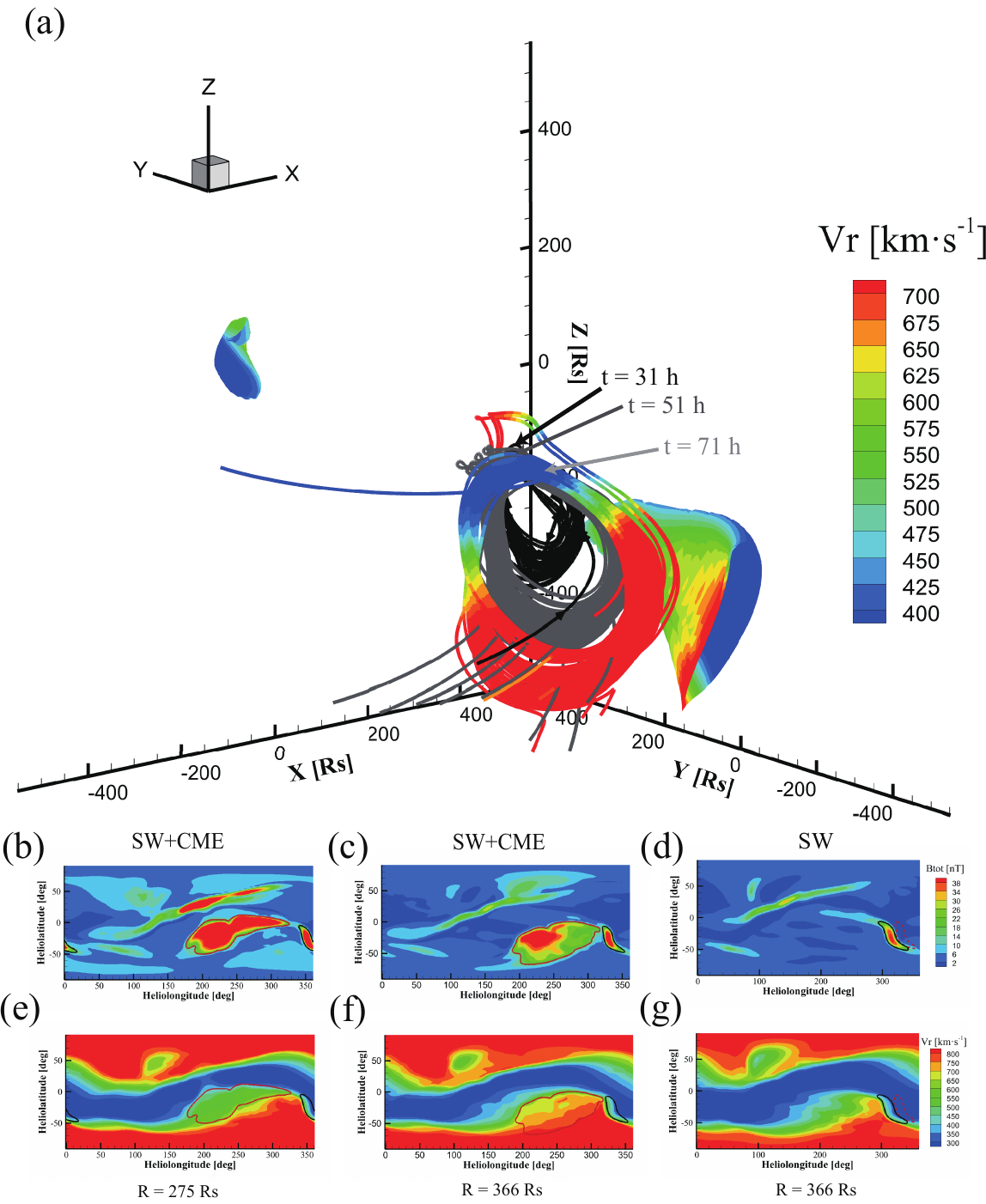}   \\
  \end{tabular}
   \end{center}
\caption{Panel (a) shows 3D magnetic flux-rope structure of the simulated CME at $t=31$ h, $t=51$ h, and $t=71$ h as well as the corotating interaction region (CIR) in the radial range between 20 and 500 Rs. Here, the colors denote the radial velocity ($V_r$). Panels (b), (c), (e), and (f) show distributions of the magnetic field strength $B_\textrm{tot}$ (b, c) and the radial velocity $V_r$ (e, f) on the surfaces at 275 Rs (b, e) and at 366 Rs (c, f), with the CME circled by the red lines and  the CIR by the black lines. Panels (d) and (g) show distributions of the $B_\textrm{tot}$ (d) and $V_r$ (g) on the surface at 366 Rs for the simulation of the background solar wind, with the CIR circled by the black lines. The area circled by the red dashed lines shows where the CIR for the simulation of the CME locates.}\label{fig6}
\end{figure}
Figure \ref{fig6} reveals again that the \text{$\alpha$} part of the CME with the speed of about 450 km s$^{-1}$ is associated with the low-speed solar wind, while the \text{$\beta$} part of the CME with the speed of about 800 km s$^{-1}$  is related to the extended fast stream. These uneven speeds across the different parts of the CME show that the CME does not move as a whole. Consequently, the CME's orientation alters.
Moreover, at the \text{$\gamma$} part of the CME, there exists a corotating interaction region (CIR) with strong magnetic fields, which may affect the CME to change its orientation due to the large pressure force \citep{Odstril1999,Isavnin2014}.
In the comparison of the positions of the CIRs without the CME (the black line) and with the CME (the red dashed line) in Figure \ref{fig6} (d) and (g), it is demonstrated that not only the CIR can deflect the CME \citep{Zhang2019, Liu2019b}, but also the CME can make the CIR deviate from its route.

\section{Summary and Discussion}
\label{sec:discuss}
In this work, we use multi-spacecraft observations and a numerical simulation to understand
the CME event on 2021 December 4, which erupts at a position at longitude $\sim 285^\circ$ and latitude $\sim -11^\circ$ in the Carrington coordinate system, and is observed as a partial halo CME from the back
side of the Sun by STEREO-A COR2 and SOHO LASCO/C3 coronagraphs. Approximately 2 days later, the CME reaches the BepiColombo spacecraft, causing significant disturbances in the surrounding magnetic field. About 4 days after the eruption, the CME arrives near Mars, where the MAVEN/Tianwen-1 detect disturbances in the magnetic field and changes in plasma parameters. The in-situ measurements shows a magnetic flux rope-like structure of the CME. By a combination of the 3D AMR-SIP-CESE MHD solar wind model with a spherical plasmoid CME model, the simulated results generally reproduce these remote sensing observations from the coronagraphs as well as in-situ measurements in the interplanetary space.

The rotation occurring during the propagation of the CME from the solar corona to Mars is emphatically investigated.
In the solar corona, the 3D distribution of the CME's flux rope-like structure as well as its symmetric axis's direction
display no significant rotation, and the CME is approximately a translational motion from 2 to 20 Rs.
However, the interplanetary propagation process (20-500 Rs) of the CME exhibits a significant rotation, with the symmetric axis's direction of
the spheromak CME deviating by approximately $40^\circ$.

It is also found that uneven motions of the CME make the CME's  magnetic flux rope-like structure rotate.
The top part of the CME lies in the area with the slow solar wind, while the bottom part is situated in fast solar wind with the open fields. Meanwhile, the right part of the flux rope is associated with a high-speed stream of the solar wind.
As the CME enters into the interplanetary space, the speed of the CME becomes close to that of the solar wind, resulting in a velocity differential of about 350 km s$^{-1}$ between the  top part  and  bottom part of the CME and about 200 km s$^{-1}$ between the left and right parts of the CME.
These uneven speed across the CME generates the  nonidentical displacements of its magnetic field lines after a time interval. That is to say, at the same time interval, the bottom part and right part of the CME's flux rope-like structure travel longer distance than the top part and right part. As a result, the axial direction of the flux rope-like structure changes and the CME rotates.
Additionally, there exists a CIR at the right part of the CME.
The interaction between the CME and the CIR not only affects the CME to change its orientation, but also
make the CIR deviate from its route.

In our study, the symmetric plasma is given at $t=0$ h within the CME rope. 
Then, the asymmetric CME expansion is formed in the corona, 
because of the interactions with the coronal background structures. The uneven plasma flows result in the interplanetary rotation of the magnetic rope.
The interactions with the ambient solar wind  enhance the rotational motions of the propagating CME.
Our study successfully reproduces the CME rotation in the interplanetary space, 
and nonetheless, it is still possible that the rotation can happen in the corona.

In the future, we plan to improve the identification of the axial direction of the CME' magnetic flux rope.
Currently, we use the average of the binormal of the magnetic field lines belonging to the CME as the  symmetry axial direction of the CME' magnetic flux rope. Although this method correctly evaluates the symmetry axial direction of the CME at the initial time, it can only qualitatively
describe the symmetry axial direction due to the distortion of the CME during the propagation in the interplanetary space. More reliable ways \citep{Rong2013} are needed to quantitatively weigh the axial direction as well as its variations with the heliocentric distance. In addition,
although a linear spheromak is in common usage to initialize the CME,
we still find some inconsistencies between the simulated CME  and the observed one, such as the magnetic field strength, the magnetic field orientation, and as on.  The CME initialization models conforming to the observations and incorporating physics \citep{Liu2019a, Singh2022} will serve as the starting point for our future simulations. The CME magnetic field is not force-free because it involves plasma (thermal and kinetic) pressures and the gravity of the Sun (at least, near the Sun). As such, the CME simulation can be also improved by considering the initiation of the CME. Last but not least,
the high-resolution modeling of CMEs with the AMR technique is our desire to understand the dynamics and structures of CMEs in greater detail. Abundant observations of CMEs from current missions \cite[e.g.][]{Vourlidas2016, Howard2020, Gan2023} and future missions \cite[e.g.][]{Wang2020, Deng2023, Gopalswamy2024}  will greatly enhance our understanding of CMEs and make numerical  results more realistic.



\section*{acknowledgments}

We would like to
thank the use of data from the SOHO and STEREO
spacecraft. We acknowledge the entire BepiColombo/MPO-MAG team for providing
data access and support. The
BepiColombo/MPO-MAG data used in the study can be
downloaded at \url{doi:10.25392 /leicester.data.22203376.v1} \citep{https://doi.org/10.25392/leicester.data.22203376.v1}.  
Tianwen-1 magnetic field data are available through the
Planet Exploration Program Scientific Data Release System
(\url{http://202.106.152.98:8081/marsdata/}) or the official website of
the MOMAG team (\url{http://space.ustc.edu.cn/dreams/tw1_momag/}). MAVEN data are available from NASA's Planetary Data System (\url{https://pds-ppi.igpp.ucla.edu/mission/MAVEN/MAVEN/}) \citep{https://doi.org/10.17189/6jmv-t325}.

This work is supported by the National Natural Science Foundation of China (Grant Nos. 42030204, 42274213, 42474216, and 42104168),  the National Key R$\&$D Program of China (grant Nos. 2022YFF0503800 and 2021YFA0718600), the B-type Strategic Priority Program of the Chinese Academy of Sciences (Grant No. XDB41000000), the Specialized Research Fund for State Key Laboratories, and the Climbing Program of NSSC (E4PD3001). The work was carried out at National Supercomputer Center in Tianjin, China, and the calculations were performed on TianHe-1 (A).


\bibliographystyle{aasjournal}



\end{document}